\documentclass[3p, 11pt]{elsarticle}
\usepackage{amsfonts,color}
\usepackage{amsmath}
\usepackage{amssymb}
\usepackage{graphicx,dsfont}
\usepackage{braket}
\usepackage[numbers]{natbib} 
\usepackage{url}
\usepackage{hyphenat}
\journal{Physics Letter B}

\newcommand{\beq}{\begin{equation}}
\newcommand{\eeq}{\end{equation}}
\newcommand{\beqa}{\begin{eqnarray}}
\newcommand{\eeqa}{\end{eqnarray}}

\begin{document}
\begin{frontmatter}
  \title{An intrinsic-state formalism for the Interacting Boson-Fermion Model with configuration mixing}

  \author[1]{Esperanza Maya-Barbecho\corref{cor1}}
  \ead{esperanza.maya@dci.uhu.es}
  
  \author[1,2]{Jos\'e-Enrique Garc\'ia-Ramos}
  \ead{enrique.ramos@dfaie.uhu.es}
  
  \address[1]{Departamento de  Ciencias Integradas y Centro de Estudios Avanzados en F\'isica, Matem\'atica y Computaci\'on, Universidad de Huelva, 21071 Huelva, Spain}
  
  \address[2]{Instituto Carlos I de F\'{\i}sica Te\'orica y Computacional, Universidad de Granada, Fuentenueva s/n, 18071 Granada, Spain}

  \cortext[cor1]{Corresponding author.}
  
  \begin{abstract}
    In this letter, we present the intrinsic state formalism of the Interacting Boson-Fermion Model incorporating both 0p-0h and 2p-2h (regular and intruder) configurations. The framework allows to deal with the unpaired fermion in either a single-j or multi-j orbital and is applicable to both axial and triaxial nuclear shapes. The formalism is applied to a schematic transitional region where regular and intruder configurations coexist and  cross. We show that the boson-fermion part of the Hamiltonian can either enhance or suppress the emergence of a type II Quantum Phase Transition.
  \end{abstract}
  
  \begin{keyword}
    Interacting Boson-Fermion Model, configuration mixing, intrinsic-state formalism
  \end{keyword}
\end{frontmatter}

\section{Introduction} 
The Interacting Boson Model (IBM) \cite{iach87} has long been established as a powerful framework for describing the collective properties of nuclei. The success of the model stems from its simplicity, rooted in its underlying algebraic structure and its close connection to the shell model \cite{Otsu1978}.

The IBM is formulated to describe nuclei with even number of protons and neutrons. However, to address nuclei containing an odd number of either protons or neutrons, the model must be extended to incorporate the interplay between bosonic (collective) and fermionic (single-particle) degrees of freedom. This extension, known as the Interacting Boson-Fermion Model (IBFM) \cite{Schol1985,Iach1991}, provides a robust framework for describing the structure of odd-mass nuclei.

The IBFM can be further extended to describe shape coexistence phenomena by incorporating multiple nuclear configurations. Within the IBM framework, this is accomplished through the configuration mixing approach (IBM-CM), which accounts for the mixing between normal (0p–0h) and intruder (2p–2h) configurations arising from excitations across closed shells \cite{duval81,duval82}. When applied to odd-mass nuclei, the corresponding extension is referred to as the Interacting Boson-Fermion Model with configuration mixing (IBFM-CM), a formalism recently introduced in \cite{Gavr22b}.

As an algebraic model, the IBFM-CM is also of particular interest from a geometric perspective. In this work, we develop an intrinsic-state formalism for the IBFM-CM based on the approach originally formulated for the IBM-CM \cite{Frank02,Frank04,Frank06}. The resulting formalism provides a framework for exploring energy surfaces, shape coexistence, and intruder band formation in odd-mass nuclei.

The intrinsic-state formalism of the IBFM was first introduced in the seminal works \cite{Levi1988,Alon1992}. In \cite{Levi1988}, the formalism was developed for a single-$j$ shell, permitting only axial shapes. In contrast, \cite{Alon1992} extended the approach to multiple-$j$ shells, where the triaxial degree of freedom began to play a significant role.
More recently, the intrinsic-state formalism has been employed in \cite{Iach2011,Petr2011} to investigate the influence of the odd fermion on the occurrence of a Quantum Phase Transition (QPT) \cite{diep80a}, concluding that the unpaired fermion can either enhance or suppress the QPT.
The formalism presented in this letter constitutes an  advancement, as it introduces an intrinsic-state approach for the IBFM-CM that is applicable to both single- and multiple-$j$ shells and accommodates both axial and triaxial shapes. A preliminary version of this formalism was previously presented in Ref.~\cite{Maya2025}.

The structure of this work is as follows. In section \ref{sec-hamiltonian}, the IBFM-CM Hamiltonian is shown. In section \ref{sec-intrinsic}, the intrinsic-state formalism is introduced. Section \ref{sec-applications} is devoted to the application of the formalism to a schematic isotopic chain, section \ref{sec-QPT} shows the relevance of the formalism under the presence of a QPT and, finally, section \ref{sec-conclu} stands for the conclusions.

\section{The Hamiltonian}
\label{sec-hamiltonian}
The IBFM \cite{Iach1991} is the natural extension of the IBM \cite{iach87}, in which the nuclear system, originally described in terms of bosons with angular momentum $L=0$ $(s)$ and $L=2$ $(d)$, is coupled to an unpaired nucleon, either a neutron or a proton. The number of bosons, $N$, corresponds to half the number of valence nucleons, excluding the unpaired one. 

The Hamiltonian of the IBFM, written in terms of boson, fermion and boson-fermion part is,
\label{sec-formalism}
\begin{equation}
  \hat{H} = \hat{H}_B + \hat{H}_F + \hat{H}_{BF}.
\end{equation}
  
The inclusion of 2p-2h excitations implies the use of a boson space $[N] \oplus [N+2]$, therefore, 
\begin{equation}
  \hat{H}=\hat{P}^{\dag}_{N}\hat{H}^N \hat{P}_{N}+
  \hat{P}^{\dag}_{N+2}\left(\hat{H}^{N+2}+\Delta^{N+2}\right)\hat{P}_{N+2}\ +\hat{V}_{\rm mix}^{N,N+2}~,
\label{ibfm_cm}
\end{equation}
where $\hat{H}^N$ stands for the regular Hamiltonian, $\hat{H}^{N+2}$ for the intruder one, $\hat{V}_{\rm mix}^{N,N+2}$ is the mixing term and $\Delta^{N+2}$ the offset between both configurations. $\hat{P}_{X}$ ($\hat{P}^\dag_{X}$) stands for the projector into the corresponding space. 

The boson Hamiltonian in each sector can be written as
\begin{equation}
  \hat{H}^i_B=\varepsilon_i \hat{n}_d+\kappa'_i  \hat{L}\cdot\hat{L}+
  \kappa_i  \hat{Q}(\chi_i)\cdot\hat{Q}(\chi_i), \label{eq:cqfhamiltonian}
\end{equation}
which is the extended consistent-Q Hamiltonian (ECQF) \cite{warner83} with $i=N,N+2$, $\hat{n}_d$ the $d$ boson number operator, being
\begin{equation}
  \hat{L}_\mu=[d^\dag\times\tilde{d}]^{(1)}_\mu ,
\label{eq:loperator}
\end{equation}
the angular momentum operator, and
\begin{equation}
  \hat{Q}_\mu(\chi_i)=[s^\dag\times\tilde{d}+ d^\dag\times
  s]^{(2)}_\mu+\chi_i[d^\dag\times\tilde{d}]^{(2)}_\mu~,\label{eq:quadrupoleop}
\end{equation}
the quadrupole operator. Where $\tilde{d}_\mu=(-1)^\mu d_{-\mu}$, $\cdot$ stands for the scalar product, and $[... \times ...]$ for the tensor product. Note that this is not the most general IBM Hamiltonian but it has been proven to properly describe a broad range of situations.

The operator $\hat{V}_{\rm mix}^{N,N+2}$ describes the mixing between
the $N$ and the $N+2$ configurations and it is defined as
\begin{equation}
  \hat{V}_{\rm B, mix}^{N,N+2}=\omega_0^{N,N+2}[s^\dag\times s^\dag + s\times s]^{(0)}+ \omega_2^{N,N+2} [d^\dag\times d^\dag+\tilde{d}\times \tilde{d}]^{(0)}.
\label{eq:vmix}
\end{equation}
Note that this term only acts on the boson sector. It is customary to assume $\omega_0^{N,N+2}=\omega_2^{N,N+2}=\omega$.

The fermion Hamiltonian only contains the single particle term because it is considered that only one fermion is present,
\begin{equation}
\hat{H}^i_F=\sum_j \epsilon_j^i \hat{n}_j
\end{equation}
where $\hat{n}_j=\sum_m a^\dagger_{j,m} a_{j,m}= -\sqrt{2j+1}(a^\dagger_j\tilde{a}_j)^{(0)}$, $\tilde{a}_{j.m}= (-1)^{j-m} a_{j,-m}$ and $\epsilon_j^i$ is the fermion single particle energy. No mixing term exist in the fermion sector. A reasonable assumption is to consider that $\epsilon_j^i =\epsilon_j$, i.e., the fermion single particle spectrum is the same in both sectors.   

Finally, for the boson-fermion part we will assume that it is composed of the monopole, the quadrupole and the exchange terms,
\begin{equation}  \hat{H}_{\mathrm{BF}}^i=\hat{H}_{\mathrm{BF}}^{\mathrm{MON},i}+\hat{H}_{\mathrm{BF}}^{\mathrm{QUAD},i}+\hat{H}_{\mathrm{BF}}^{\mathrm{EXC},i},
\end{equation}
where,
\begin{equation}
  \begin{aligned}
    & H_{\mathrm{BF}}^{\mathrm{MON},i}=\sum_j A_j^i\left[\left[d^{\dagger} \times \tilde{d}\right]^{(0)} \times\left[a_j^{\dagger} \times \tilde{a}_j\right]^{(0)}\right]_0^{(0)}, \\
    & H_{\mathrm{BF}}^{\mathrm{QUAD},i}=\sum_{j j^{\prime}} \Gamma_{j j^{\prime}}^i\left[\hat{Q}(\chi_i) \times \left[a_j^{\dagger} \times \tilde{a}_{j^{\prime}}\right]^{(2)}\right]_0^{(0)}, \\
    & H_{\mathrm{BF}}^{\mathrm{EXC},i}= \sum_{j j^{\prime} j^{\prime \prime}} \Lambda_{j j^{\prime}}^{j^{\prime \prime},i}:\left[\left[d^{\dagger} \times \tilde{a}_j\right]^{\left(j^{\prime \prime}\right)} \times\left[\tilde{d} \times a_{j^{\prime}}^{\dagger}\right]^{\left(j^{\prime \prime}\right)}\right]_0^{(0)}:
\end{aligned}
\end{equation}
and $:$ stands for normal order.

The boson-fermion mixing term is assumed to have a similar structure that the boson one and the coefficient of both components are taken equal,
\begin{equation}
  \hat{V}_{\rm BF, mix}^{N,N+2}=\sum_j\hat{n}_j\omega_j\big([s^\dag\times s^\dag + s\times s]^{(0)} + [d^\dag\times d^\dag+\tilde{d}\times \tilde{d}]^{(0)}\big),
\label{eq:vmix_fermion}
\end{equation}
but in most cases, $\omega_j$ can be considered independent on $j$ and thus its effect can be absorbed in the boson mixing term. As a result, the mixing term remains diagonal in the fermion part of the wave function.

\section{The intrinsic state formalism}
\label{sec-intrinsic}
The proposed intrinsic state formalism of the IBFM-CM is constructed in terms of the original intrinsic state of the IBM \cite{gino80,diep80a,diep80b},
\begin{equation}
\label{GS}
|N; \beta,\gamma   \rangle = {1 \over \sqrt{N!}}
\Bigg({1 \over \sqrt{1+\beta^2}} \Big (s^\dagger + \beta
\cos     \gamma          \,d^\dagger_0\\          
+{1\over\sqrt{2}}\beta
\sin\gamma\,(d^\dagger_2+d^\dagger_{-2})\Big) \Bigg)^N | 0 \rangle ,
\end{equation}
but adding a fermion that can be in any of the available shells, therefore, in order to take into account the fermion degree of freedom, a matrix form should be introduced,
\begin{equation}
\label{E_ibfm}
\begin{aligned}
    {\cal H}_{IBFM}(N,\beta,\gamma) &= 
    \sum_{j_1, m_1, j_2, m_2} {\cal M}(N,\beta,\gamma)_{j_1, m_1, j_2, m_2} a^\dagger_{j_1,m_1} a_{j_2,m_2}\\
    &=
    \sum_{j, m} \big(E_{B}(N,\beta,\gamma)+ \epsilon_{j}\big ) a^\dagger_{j,m} a_{j,m} \\
    &+\sum_{j_1, m_1, j_2, m_2} E_{BF}(N, \beta, \gamma)_{j_1, m_1, j_2, m_2}
    \times\frac{a^\dagger_{j_1,m_1} a_{j_2,m_2}+ a^\dagger_{j_2,m_2} a_{j_1,m_1}}{1+\delta_{j_1,j_2}\delta_{m_1,m_2}}.  
\end{aligned} 
\end{equation}
$\epsilon_j$ are the fermion single quasi-particle energies, $E_{B}(N,\beta,\gamma)$ stands for the matrix element of $\hat{H}_B$ with the intrinsic state (\ref{GS}) (see \cite{gino80}), and $E_{BF}(N, \beta, \gamma)_{j_1,m_1, j_2, m_2}$ is the matrix element of $\hat{H}_{BF}$ (see \cite{Petr2011} although only for a single j shell). 
The implicit basis used to define the Hamiltonian is the direct product of the intrinsic state for bosons and the fermion part,
\begin{equation}
    | N;\beta,\gamma; j m\rangle = |N; \beta,\gamma   \rangle \otimes | j m \rangle.
\end{equation}

$E_{B}(N,\beta,\gamma)$ can be written as,
\begin{equation}
\label{E_B}
\begin{aligned}    
E_{B}(N,\beta,\gamma)&=\frac{N \beta^2}{1+\beta^2}(\varepsilon+6\kappa') +
\frac{N}{1+\beta^2}\,\kappa\,(5+(1+\chi^2)\beta^2)\\
&+\frac{N(N-1)}{(1+\beta^2)^2}\,\kappa\, 
\Big(\frac{2}{7}\chi^2\beta^4-4\sqrt{\frac{2}{7}}
\chi\beta^3\cos{3\gamma}+4\beta^2\Big).
\end{aligned}
\end{equation}

Regarding $E_{BF}(N, \beta, \gamma)_{j_1, m_1, j_2, m_2}$, the different contributions are,
\begin{equation}
    \begin{aligned}
       E_{BF}^{\rm{MON}}(N, \beta, \gamma)_{j_1, m_1, j_2, m_2}=-N A_{j_1}/\sqrt{5(2 j_1+1)} \delta_{j_1,j_2}\delta_{m_1,m_2} \frac{\beta^2}{1+\beta^2}, 
    \end{aligned}
\end{equation}

\begin{equation}
    \begin{aligned}
       E_{BF}^{\rm{QUAD}}(N, \beta, \gamma)_{j_1, m_1, j_2, m_2} = N \frac{\Gamma_{j_1 j_2}}{\sqrt{5}} 
        \sum_M(-1)^{j_2+m_2-M} q_{2M}(\beta,\gamma)
        \langle j_1, m_1; j_2, -m_2 | 2 -M\rangle, 
    \end{aligned}
\end{equation}
where 
\begin{equation}
\begin{aligned}
& q_{20}(\beta, \gamma)=\frac{1}{1+\beta^2}\left(2 \beta \cos \gamma-\sqrt{\frac{2}{7}} \beta^2 \chi \cos 2 \gamma\right), \\
& q_{2\pm 2}(\beta, \gamma)=\frac{1}{1+\beta^2}\left(\sqrt{2} \beta \sin \gamma+\sqrt{\frac{1}{7}} \chi \beta^2 \sin 2 \gamma\right),\\
& q_{2\pm 1}(\beta, \gamma)=0,
\end{aligned}
\end{equation}
and
\begin{equation}
    \begin{aligned}
       E_{BF}^{\rm{EXC}}(N, \beta, \gamma)_{j_1, m_1, j_2, m_2} &= 
       N\; \sum_{ j'' m''}\Lambda^{j''}_{j_1 j_2}\dfrac{(-1)^{j_1+j''}}{\sqrt{2j''+1}} \eta_{ m''+m_1} \eta_{ m''+m_2}\\
       &\times\langle 2, m''+m_1; j_1, -m_1 | j'' m''\rangle \langle 2, -m_2-m''; j_2, m_2 | j''-m''\rangle,
    \end{aligned}
\end{equation}
where
\begin{equation}
    \begin{aligned}
        \eta_{0} = \dfrac{\beta\cos\gamma}{\sqrt{1+\beta^2}},\qquad
        \eta_{1} = \eta_{-1} = 0,\qquad 
        \eta_{2} = \eta_{-2} = \dfrac{\beta\sin\gamma}{\sqrt{2(1+\beta^2)}}
    \end{aligned}
\end{equation}
and $\langle ...| ...\rangle$ stand for the Clebsh-Gordan coefficients. Hence, solving the problem requires diagonalizing a matrix of dimension $\sum_j (2 j+1)$. However, since the Hamiltonian only connects states with $\Delta m=\pm 2$, the matrix can be decomposed into two equivalent blocks, each of dimension $1/2\sum_j (2 j+1)$. One block includes states with $m=j, j-2, ..., -(j-1)$, while the other contains the corresponding complementary values \cite{Petr2011}.

Former expressions can be reduced to a more compact form in the case of a single $j$ shell, where $\Gamma_{jj'}\rightarrow \Gamma_j $ and $\Lambda_{jj'}^{j''}\rightarrow \Lambda^j$ and  axial symmetry is considered (see \cite{Petr2011}). Indeed, the matrix takes a diagonal form in terms of the projection $K$,
\begin{equation}
\begin{aligned}
E(\beta)_{j,K}&= E_{B}(N,\beta,0) +\epsilon_j -A_j\frac{N}{\sqrt{5(2j+1)}}\frac{\beta^2}{1+\beta^2}\\
&-\Gamma_j \frac{N}{1+\beta^2} (2\beta -\beta^2 \chi \sqrt{2 / 7}) [3 K^2-j(j+1)] P_j\\
&-\Lambda_j N \sqrt{2j+1} \frac{\beta^2}{1+\beta^2} [3 K^2-j(j+1)]^2 P_j^2,
\end{aligned}
\end{equation}
where $P_j$ stands for,
\begin{equation}
    P_j \equiv \left[(2j-1)j(2j+1)(j+1)(2j+3)\right]^{-1/2}.
\end{equation}

When introducing the regular and the intruder configurations, it is needed to enlarge the considered matrix in a similar way to the case of bosons as in \cite{Frank02,Frank06},
\begin{equation}
\label{H_CM}
{\cal H}_{IBFM}^{CM}(N,\beta,\gamma)=\left (
\begin{array}{@{\hspace{0.2em}}c@{\hspace{0.3em}}c@{\hspace{0.2em}}}
{\cal M}(N,\beta,\gamma)& \Omega_{BF}(\beta)\\
\Omega_{BF}(\beta)& {\cal M}(N+2,\beta,\gamma)
\end{array}
\right ),
\end{equation}
where ${\cal M}(N,\beta,\gamma)$ and ${\cal M}(N+2,\beta,\gamma)$ correspond to the matrices of the regular and the intruder Hamiltonian and $\Omega_{BF}(\beta)$ is the matrix form of the mixing operator $V_{mix}(N, \beta)$ written as
\begin{equation}
\label{V_mix}
   V_{mix}(N,\beta)=\sum_{j_1, m_1, j_2, m_2} \Omega_{BF}(N,\beta)_{j_1, m_1, j_2, m_2}\; a^\dagger_{j_1,m_1} a_{j_2,m_2}.  
\end{equation}
The expression of $V_{mix}(N,\beta)$, assuming equal weight for the $s$ and $d$ boson parts, can be written as,
\begin{equation}
    \Omega_{BF}(N,\beta)_{j_1, m_1, j_2, m_2}=\delta_{j_1,j_2}\delta_{m_1,m_2}(\omega+\omega_{j_1})\frac{\sqrt{(N+2)(N+1)}}{1+\beta^2}\left(1+\frac{\beta^2}{\sqrt{5}}\right),
\end{equation}
but considering no dependence of $\omega_j$ on $j$, its value can be absorbed in $\omega$, such that $\omega+\omega_{j}\rightarrow\omega$. Therefore, $\Omega_{BF}(N,\beta)$ is a diagonal matrix proportional to the identity.

Consequently, the new formalism introduces several important advancements: it incorporates all three components of the boson-fermion interaction, accounts for multiple $j$ values, includes both $N$ and $N+2$ configurations, and considers the triaxial degree of freedom. These features represent a step forward in the development of the boson-fermion intrinsic-state formalism within the IBFM framework \cite{Levi1988, Alon1992, Iach2011, Petr2011}.

As it was already mentioned, the matrix (\ref{H_CM}) has a dimension of $2\sum_j (2j+1)$, considering the $N$ and $N+2$ sectors, but it can be split in two equivalent blocks and its diagonalization provides $\sum_j (2j+1)$ eigenvalues that depend on $\beta$ and $\gamma$ and that are doubly degenerated.
For each of them, one has to obtain the value of the equilibrium deformation parameters $(\beta_0, \gamma_0)$, minimizing the corresponding energy surfaces \cite{Petr2011}. Note that the equilibrium parameters will be different, in principle, for the different energy surfaces. 

\begin{figure}[hbt]
\centering\includegraphics[width=0.45\linewidth]{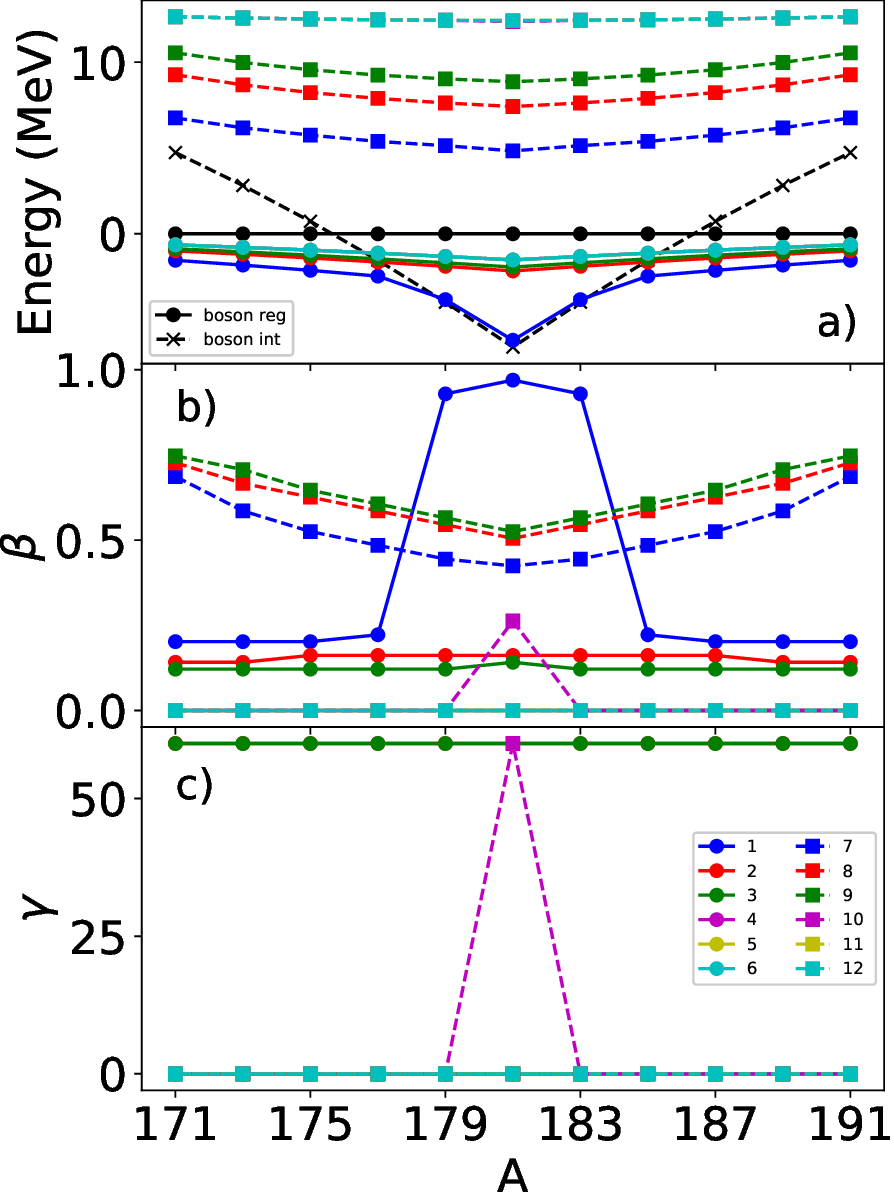}
\caption{Equilibrium value of energies and $\beta$ and $\gamma$ deformation parameters for the schematic isotope chain. a) Unperturbed boson energies (black lines) and fermion energies of the different orbits. b) Equilibrium value of $\beta$ for the fermion orbits. c) Equilibrium value of $\gamma$ for the fermion orbits. Full lines with full circles for the six lower levels, dashed lines with full squares for the six upper ones. Colors stand to distinguish the order of the levels.
\label{fig-equilibrium}}
\end{figure}

\section{Application to a crossing of configurations}
\label{sec-applications}
As an application of the new formalism, we consider a scenario inspired by Ref.~\cite{harder97}, in which a spherical configuration coexists with a deformed one in even-even Pt isotopes. Specifically, we examine a hypothetical even-even isotope chain ranging from  $A=172$ to $A=192$, with  mid-shell located at $A=182$. For the regular part of the Hamiltonian, we set $\varepsilon^N= 1$ MeV, while all other regular parameters are set to zero. In the intruder configuration, the parameters are $\kappa^{N+2}=-0.044$ MeV, $\chi^{N+2}=-\sqrt{7}/2$  with the remaining intruder parameters also set to zero. The mixing interaction is characterized by $\omega=0.15$ MeV, and the energy offset for the intruder configuration is  $\Delta^{N+2}=14.2$ MeV.  
Accordingly, the regular configuration corresponds to the vibrational U(5) limit, while the intruder configuration corresponds to the SU(3) rotor structure. The number of bosons is assumed to evolve from $N=8$ to $N=13$ at mid-shell and then return to $N=8$ at the end of the chain.  Thus, we assign $N=8$ to $A=172$, $N=13$ to the mid-shell nucleus at $A=182$, and again $N=8$ to $A=192$.  The value of $\Delta^{N+2}$ has been chosen to induce the crossing of the boson configurations between $A=176$ and $A=178$, i.e., $N=10$ and $11$. Therefore, a type II QPT emerges at this point (see Section \ref{sec-QPT}), as observed in Fig.~\ref{fig-equilibrium}a. 
The odd fermion added to the system will be a proton hole, hence, the hypothetical isotope chain  correspond to Ir and therefore a unit should be subtracted to $A$.
\begin{figure}[hbt]
\centering\includegraphics[width=0.5\linewidth]{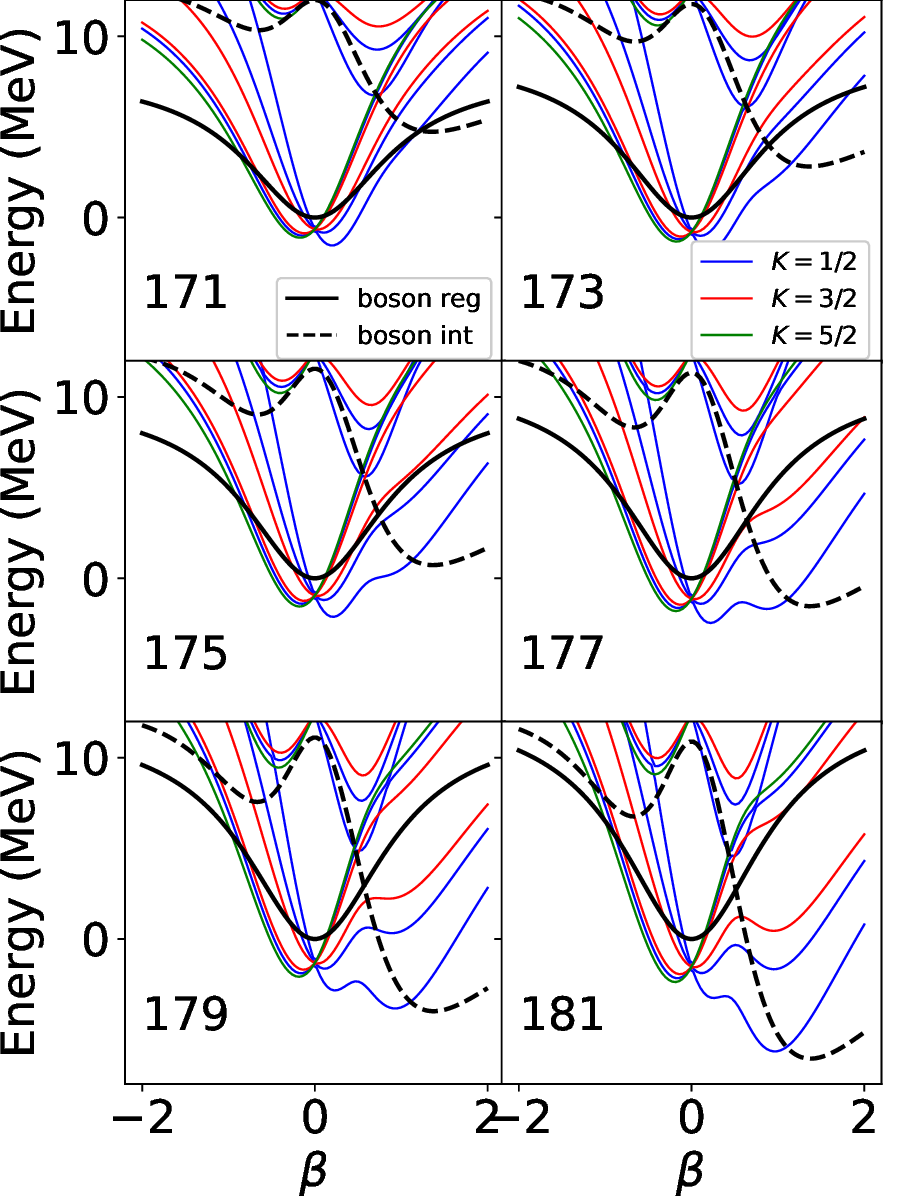}
\caption{Axial energy curves for the fermion orbits over the isotope chain. The fermion lines are marked with its $K$ value. The thick black lines correspond to the unperturbed boson energy curves, full for the regular and dashed lines for the intruder configuration.}
\label{fig-axial}
\end{figure}

The effect of the different boson-fermion terms in a single-j shell and a single particle-hole configuration was previously studied in \cite{Petr2011}. 
In this work, we consider a specific case involving two configurations using a set of parameters that maximize the effect of the different boson-fermion terms in a multi-j situation, with $j=1/2, 3/2, 5/2$. A more comprehensive analysis will be presented in a future publication \cite{Maya2025u}. Specifically, we adopt the following set of boson-fermion parameters,
\begin{equation}
    \begin{aligned}
    \epsilon_j^N=\epsilon_j^{N+2}&=0\, \text{MeV},\\   
        A_0^N=A_0^{N+2}&= 1.0\, \text{MeV}, \\
        \Gamma_0^N=\Gamma_0^{N+2}&=-5.0 \, \text{MeV},\\
        \Lambda_0^{N}=\Lambda_0^{N+2}&=1.0  \, \text{MeV}, \\
        \chi^{N}=-\chi^{N+2}&=\sqrt{7}/2, \\
    \end{aligned}
\end{equation}
Note that we closely follow the microscopic interpretation of the IBFM proposed in \cite{Schol1985}.
\begin{equation}
\begin{aligned}
A_j & =-\sqrt{5(2 j+1)} A_0, \\
\Gamma_{j j^{\prime}} & =\sqrt{5} \gamma_{j j^{\prime}} \Gamma_0, \\
\Lambda_{j j^{\prime}}^{ j^{\prime \prime}} & =-2 \sqrt{\frac{5}{2 j^{\prime \prime}+1}} \beta_{j j^{\prime \prime}} \beta_{j^{\prime} j^{\prime \prime}} \Lambda_0,
\end{aligned}
\end{equation}
where
\begin{equation}
    \begin{aligned}
\gamma_{j j^{\prime}} & =\left(u_j u_{j^{\prime}}-v_j v_{j^{\prime}}\right) Q_{j j^{\prime}}, \\
\beta_{j j^{\prime}} & =\left(u_j v_{j^{\prime}}+v_j u_{j^{\prime}}\right) Q_{j j^{\prime}}, \\
Q_{j j^{\prime}} & =\left\langle j|| Y^{(2)} \| j^{\prime}\right\rangle,
\end{aligned}
\end{equation}
being $Y^{(2)}$ the spherical harmonic. For our schematic calculation, we impose $v_j=v$ ($v^2+u^2=1$), with $N_p = \sum_j (2j+1) v_j^2$ and $N_p=7$.
\begin{figure}[hbt]
\centering\includegraphics[width=0.6\linewidth]{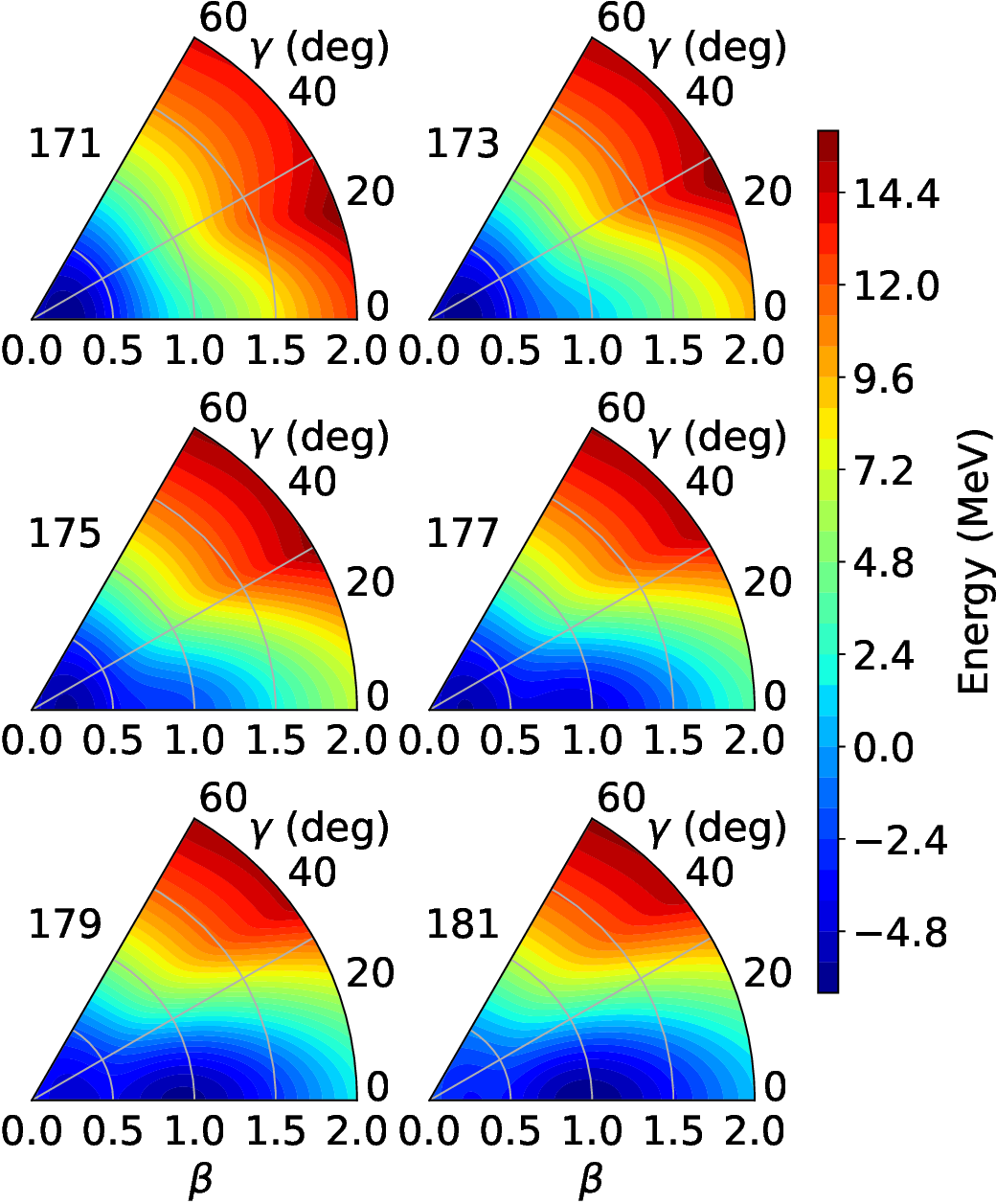}
\caption{Triaxial energy surface for the first orbit along the isotope chain.}
\label{fig-triaxial}
\end{figure}

In Fig.~\ref{fig-equilibrium}, the energy of the single particle energies (panel a) are presented, corresponding to the equilibrium values of $\beta$ and $\gamma$ shown in panels b) and c), respectively. Note that $12$ levels appear, although they are doubly degenerated. Each energy level corresponds to a distinct equilibrium value of $\beta$ and $\gamma$; therefore, different states can exhibit different shapes or degrees of deformation. The key observation in the trend is that only the lowest energy level undergoes a transition from a regular (approximately spherical) to an intruder (more deformed) configuration. In contrast, the other low-lying states maintain a predominantly spherical character throughout. Additionally, the six higher-lying states (represented by dashed lines) are primarily associated with the intruder configuration and exhibit a prolate deformation.

The equilibrium deformation values and the corresponding energies alone do not fully capture the complexity of the energy surface topology. Therefore, in Fig.~\ref{fig-axial}, we present the axial fermion energy curves across the entire isotope chain, alongside the unperturbed boson energy curves used as a reference (note that the second half of the shell is symmetric with respect to the first). 
The levels are labeled with the $K$ quantum number, which is preserved for the case of $\gamma=0$.
By construction, the regular states exhibit a spherical shape, while the intruder states are associated with deformation.
The deformation observed in the lower states arises from the lowering of the intruder levels toward mid-shell, as well as from the interaction between the regular and intruder configurations. Notably, secondary minima appear in several of the energy curves; these minima can become nearly degenerate and may even exchange order as one moves along the isotope chain.

To gain a comprehensive understanding of the evolution of the single-particle levels, it is more informative to examine the behavior of the energy surface in the $\beta-\gamma$ plane. Here, we focus on the evolution of the lowest energy surface, shown in Fig.~\ref{fig-triaxial}, although the remaining surfaces can be readily computed and will be presented in a forthcoming publication \cite{Maya2025u}. The results reveal a smooth transition from a spherical shape in nuclei with $A=171–173$ to a deformed structure for $A=179–181$. Notably, for intermediate nuclei ($A=175–177$), two coexisting minima emerge (see also Fig.~\ref{fig-axial}). The presence of such coexisting minima is a distinctive feature of the present formalism. However, the specific boson-fermion Hamiltonian employed in this study tends to suppress the development of the second minimum and delays the onset of deformation. It is important to emphasize that this behavior is not universal, but rather depends on the particular choice of boson-fermion interaction.

\section{On Quantum Phase Transitions}
\label{sec-QPT}
The concept of QPT has been extensively explored in nuclear physics within the frameworks of both the IBM and the IBFM, with and without configuration mixing. In the case of the IBM with configuration mixing (IBM-CM), several seminal studies \cite{Frank06, Hell07} have demonstrated that the model's phase diagram is particularly rich and complex, allowing for the emergence of both first- and second-order QPTs.

Notably, in the presence of two configurations, a first-order QPT can occur even when $\chi=0$, a situation that is not possible in the IBM with a single configuration.  In such cases, the structure of the phase diagram becomes highly sensitive to the value of $\Delta^{N+2}$ and, for sufficiently large values, the QPT may no longer appear. The origin of the first-order transition lies in the topology of the ground-state energy surface, which, according to Catastrophe Theory \cite{Gilm1981}, has a ``butterfly'' germ characterized by a $\beta^6$ dependence. 

In the case of the IBFM with a single configuration, Refs.~\cite{Iach2011, Petr2011} showed that, for a Hamiltonian depending continuously on a single parameter, the occurrence of the phase transition can be either enhanced or suppressed depending on the $m$-projection of the fermion level.  In this context, it is interesting to analyze the effect of the different boson-fermion interactions. On the one hand, the monopole and exchange terms generate a contribution proportional to $\frac{\beta^2}{1+\beta^2}$, whose effect can be effectively absorbed into the one-body boson term. On the other hand, the quadrupole term introduces not only this same $\frac{\beta^2}{1+\beta^2}$ dependence, but also an additional$\frac{\beta}{1+\beta^2}$ term. This latter contribution is crucial, as it can give rise to a first-order QPT, even in the case of a single configuration with $\chi=0$ in the boson part of the Hamiltonian.
\begin{figure}[hbt]
\centering\includegraphics[width=0.8\linewidth]{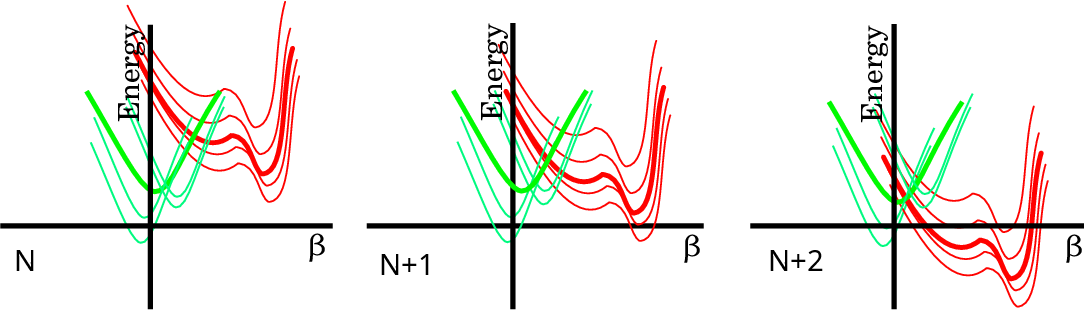}
\caption{Schematic evolution of unperturbed regular (spherical, green lines) and intruder (deformed, red lines) configurations for the IBFM. The thick lines correspond to the boson energy surfaces while the thin ones to the fermion levels.}
\label{fig-schematic}
\end{figure}

In general, determining the phase diagram of the IBFM with configuration mixing (IBFM-CM) is a highly complex task, requiring both analytical approaches based on Catastrophe Theory and detailed numerical analyses. However, in this work, we present qualitative arguments to gain insight into the influence of the odd fermion and the interplay between the two particle-hole configurations.

Specifically, we aim to interpret the behavior observed in Fig.\ref{fig-axial}, where the low-lying states exhibit increasing deformation as the system approaches mid-shell, despite the Hamiltonian remaining constant. This trend can be attributed to the increasing number of bosons along the isotopic chain, which enhances the binding energy of the deformed configuration. As a result, the spherical and deformed configurations eventually cross and become nearly degenerate. This crossing mechanism is schematically illustrated in Fig.\ref{fig-schematic}.

The presence of the odd fermion modifies the crossing behavior in several ways. First, the boson energy levels are split according to the $K$-projection of the fermion (within the axial approximation), which can either raise or lower the energy depending on the specific boson-fermion interaction. Second, the deformation associated with each configuration is slightly modified by the coupling to the fermion. Furthermore, the mixing between the regular and intruder configurations induces level repulsion, which can affect the relative ordering of states.

Taken together, this schematic picture suggests that a type II QPT can be induced in the IBFM-CM framework, as previously proposed in Refs.~\cite{Gavr22b, Gavr2023, Gavr2025}. Importantly, the precise mass number at which the QPT occurs is shifted relative to the purely bosonic case, due to the influence of the odd fermion. Given the coexistence of multiple minima in the energy surface, the resulting QPT is expected to be of first order.

\section{Conclusions}
\label{sec-conclu}
In this letter, we have presented the intrinsic state formalism for the IBFM with configuration mixing, applicable to both single- or multiple-$j$ shells, axial and triaxial shapes. It has been applied to a multiple-$j$ shell case, using an IBFM Hamiltonian that includes U(5) and SU(3) boson terms for the regular and the intruder configurations, respectively. The considered boson-fermion interaction contains the monopole, the quadrupole and the exchange components. Axial and $\beta-\gamma$ fermion energy surfaces are calculated along with the corresponding equilibrium values. In the presented example, a type II QPT and clear signatures of shape coexistence are observed.

While completing this work, we became aware of the publication of the similar work \cite{Levi2025}.

\section{Acknowledgments}
The authors thank J.M.~Arias and P.~Van Isacker for useful discussions. This work was partially supported by grant  PID2022-136228NB-C21 funded by MCIN/AEI/10.13039/50110001103 and ``ERDF A way of making Europe''. EMB thanks economical support from grant AST22-00001-X funded by EU "NextGenerationEU", by MCIN/AEI/10.13039/50110001103, by  Consejer\'ia de Universidad, Investigaci\'on e Innovaci\'on de la Junta de Andaluc\'ia and by University of Huelva. Resources supporting this work were provided by the CEAFMC and Universidad de Huelva High Performance Computer (HPC@UHU) funded by ERDF/MI\-NE\-CO project UNHU-15CE-2848.


\end{document}